# FABRICACIÓN DE UN MAGNETRÓN SPUTTERING PARA DEPOSITO DE PELÍCULAS NANOMÉTRICAS MAGNÉTICAS

**D. Ley Domínguez[1], César L. Ordóñez-Romero[1], G. Pirruccio[1], F. Ascencio-Aguirre[2], Ana Bobadilla[2]**

[1]Instituto de física, Universidad Nacional Autónoma de México, CU, 04510 D.F., México
[2]Instituto de Investigaciones en Materiales, Universidad Nacional Autónoma de México, CU, 04510 D.F., México

**Resumen:** En este trabajo se presenta el desarrollo de instrumentación científica para la fabricación de películas delgadas ferromagnéticas, por la técnica de pulverización catódica (sputtering), para el uso de blancos de 2 pulgadas de diámetro. Se depositaron películas delgadas utilizando como material ferromagnético la aleación de Permalloy ($Ni_{80}Fe_{20}$) a temperatura ambiente, sobre sustratos de Si (001). Los espesores de las películas se midieron con perfilometria y se calculó una tasa de depósito para esta aleación de 16.2 nm/min. Microscopia electrónica de barrido mostró una formación de película continua y una composición química similar a la del blanco.

**Palabras clave:** Magnetrón de pulverización catódica, Películas delgadas, tasa de depósito

**Abstract:** This paper presents the development of scientific instrumentation for the fabrication of ferromagnetic thin films, by sputtering technique, for the use of 2-inch-diameter targets. Thin films were deposited using Permalloy alloy ($Ni_{80}Fe_{20}$) as ferromagnetic material at room temperature on Si (001) substrates. The film thicknesses were measured with profilometry and a deposition rate for this alloy of 16.2 nm/min was calculated. Scanning electron microscopy showed a continuous film formation and a chemical composition similar to the target.

## 1. Introducción

La pulverización catódica (sputtering) es una técnica para fabricar películas delgadas mediante la deposición física de vapor, fue descubierta y reportada hace más de 100 años por Grove (Grove, 1852) en 1852 y de forma independiente por Plücker (Plücker, 1858) en 1858. Ya para 1877 la pulverización catódica estaba siendo utilizada para la fabricación de espejos. Por 1930 esta técnica fue utilizada para recubrimientos en patrones de fonógrafos (Behrisch R. K. W., 1991).

En la pulverización catódica la superficie de un sólido es bombardeada con iones energéticos y los átomos de esta superficie son dispersados debido a colisiones entre los átomos de la superficie y las partículas energéticas. Estos átomos removidos son depositados en un sustrato formando una película delgada.

En 1935 Penning, fue el primero en estudiar la técnica de pulverización catódica colocando un campo magnético en el área del plasma, formando un magnetrón. En un sistema de pulverización catódica de magnetrón, un campo magnético es superpuesto sobre el cátodo de forma paralela a la superficie del cátodo. Los electrones en la descarga luminiscente muestran un movimiento en forma de cicloide donde el centro de sus orbitas se desplaza en una dirección E x B, donde E y B son el campo eléctrico en la descarga y el campo magnético transversal superpuesto, respectivamente. El campo magnético está orientado de tal manera que los electrones desviados formen un circuito cerrado. Este efecto de atrapamiento de electrones aumenta la tasa de colisión entre los electrones y las moléculas del gas de pulverización catódica, con esto se consiguió bajar la presión de gas de depósito en un factor de 10 e incrementar la tasa de depósito de las películas (Wasa K. M. K., 2004), (Wasa K. I. K., 2012).

Sin embargo no fue hasta la década de 1950s que un mejor entendimiento de esta técnica y la demanda de películas delgadas de gran calidad de diferentes materiales, principalmente estimulada por avances tecnológicos de microelectrónica y semiconductores impulsaron nuevos desarrollos de esta técnica (Maissel, 1970). A principio de 1960s la técnica de pulverización catódica de magnetrón fue reconsiderada como un proceso atractivo para el depósito de películas delgadas. En 1967 Wasa, invento un prototipo de magnetrón plano, y en 1974 Capin, diseño un sistema practico de magnetrón con una bobina solenoide, similar a los utilizados hoy en día.

En el presente, la pulverización catódica permite un buen control de composición, dimensiones y una gran flexibilidad en los tipos de materiales que pueden depositarse. Esta técnica se utiliza para muchas aplicaciones y se ha convertido en un proceso indispensable en la nanotecnología moderna y la física. Las películas delgadas obtenidas por pulverización catódica son utilizadas para la fabricación de nanoestructuras, nanodispositivos, Sistemas-Micro-Electro-Mecánicos (MEMS), (Bhushan, 2017) así como para la fabricación de muestras de numerosos materiales para la investigación de diversas propiedades. En la industria, las películas delgadas obtenidas por pulverización catódica son utilizadas en dispositivos electrónicos, recubrimientos ópticos, revestimientos duros para instrumentos, dispositivos magnéticos de almacenamiento de datos y accesorios decorativos (Behrisch R. W. E., 2007).

En este artículo se presenta la fabricación de un magnetrón de pulverización catódica para blancos ferromagnéticos de 2 pulgadas, así como la caracterización de la tasa de depósito para películas de permalloy sobre sustratos de Si (001), morfología y análisis elemental semicuantitativo con microscopia electrónica de barrido.

## 2. Diseño y experimentación

El diseño del magnetrón se dividió en tres partes; núcleo, carcasa y tapa. La figura 1 muestra el esquema del magnetrón en donde se observan los diferentes componentes y materiales que lo conforman. El núcleo del magnetrón, mostrado en la figura 2, es unas de las partes más importantes, como se comentó en la introducción para que el magnetrón funcione se requiere de un campo eléctrico y un campo magnético, por lo tanto al cerrarse el circuito eléctrico se generara calor, por esta razón el núcleo cuenta con un soporte de cobre mostrado en la figura 2 b), ya que su buena conductividad térmica se utiliza para funcionar como intercambiador de calor con el refrigerante y su buena conductividad eléctrica es ideal para formar parte del electrodo. En esta parte se ubican los súper imanes de NdFeB que generan el campo magnético del magnetrón.

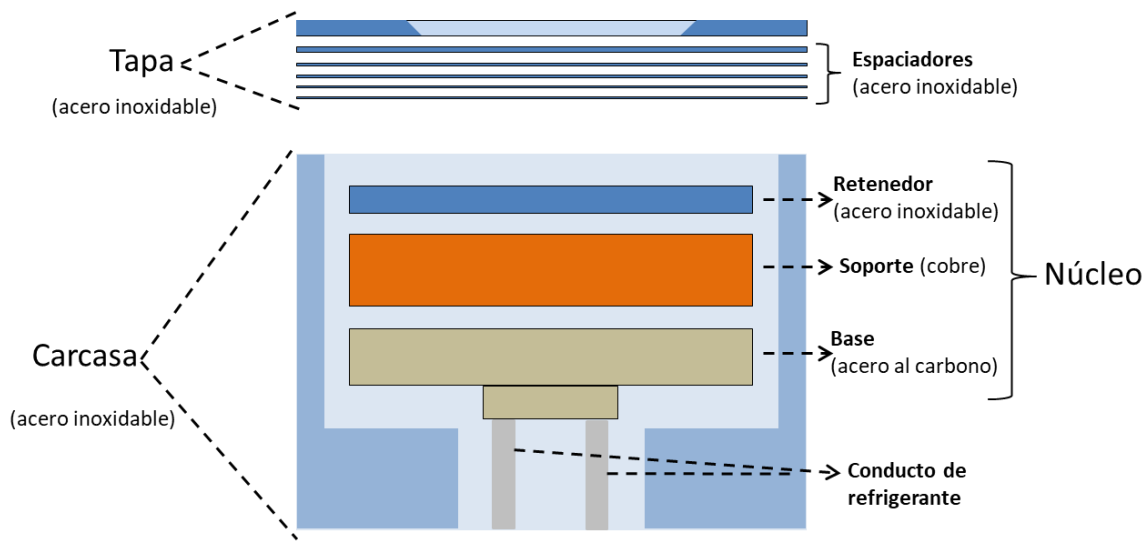

*Figura 1. Esquema del diseño del magnetrón de pulverización catódica.*

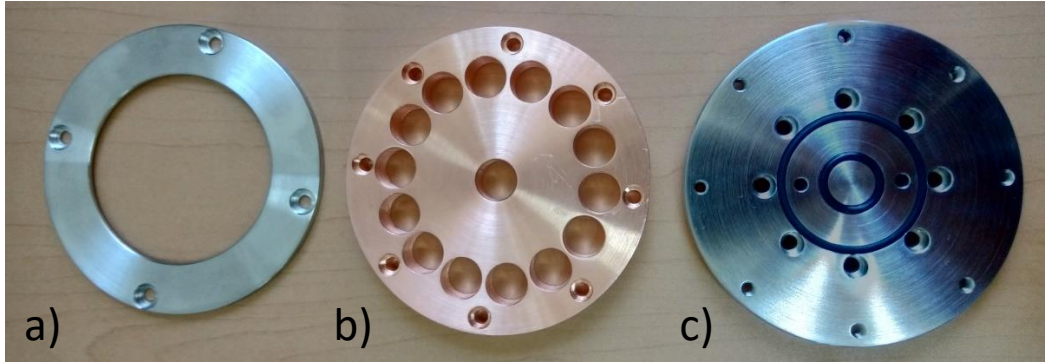

*Figura 2. a) Tapa, b) soporte y c) base, que conforman el núcleo del magnetrón.*

El núcleo también cuenta con una base ferromagnética de acero al carbono mostrada en la figura 2 c), utilizada para cerrar las líneas del campo magnético, esta base también cuenta con juntas tóricas para sellar la zona del intercambiador de calor. La tapa mostrada en la figura 2 a) se utiliza para sujetar el blanco, es de acero inoxidable no magnética para no interferir con las líneas de campo magnético del magnetrón. La carcasa mostrada en la figura 3 es parte del otro electrodo y fue fabricada de acero inoxidable no magnético para no interferir con el campo magnético. La tapa mostrada en la figura 4 también fabricada de acero inoxidable no magnético cuenta con espaciadores de diferente grosor para modificar la distancia entre los electrodos dependiendo del espesor de los blancos. La figura 5 muestra el magnetrón sin la tapa donde se pueden observar los súper imanes de NdFeB montados en el soporte de cobre. En la figura 6 se muestra el magnetrón ensamblado en el brazo de la cámara de vacío.

Para la fabricación de las películas delgadas, el magnetrón fue montado dentro de una cámara de vacío con una presión base de $5 \times 10^{-6}$ Torr, después fue presurizada con flujo de argón con una pureza de 99.998% a $3.0 \times 10^{-3}$ Torr durante el depósito de las películas. Para el depósito se utilizó una fuente de RF fijada a una potencia de 40W, se utilizó un blanco de material ferromagnético de 2 pulgadas de diámetro de Permalloy ($Ni_{80}Fe_{20}$) con una pureza de 99.95%. Las películas fueron depositadas a una distancia de 5cm del magnetrón, sobre sustratos de silicio monocristalino con orientación (001), un espesor de 525 µm, un área de 5 mm x 7 mm y previamente lavados con ultrasonido. Las películas se fabricaron variando el tiempo de depósito y su tasa de depósito fue caracterizada con un perfilómetro Veeco Dektak 150. La morfología y composición química fue caracterizada por microscopia electrónica de barrido con un microscopio JEOL JSM-7600F.

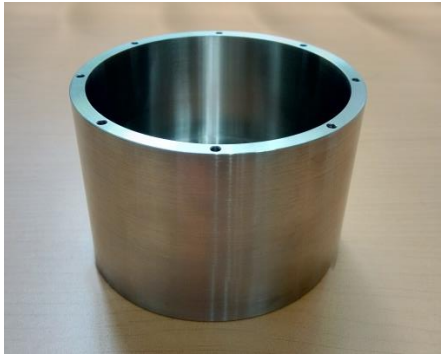
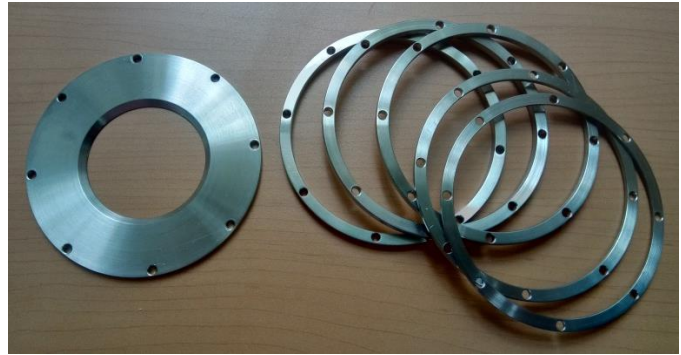

*Figura 3. Carcasa del magnetrón.*      *Figura 4. Tapa y espaciadores del magnetrón.*

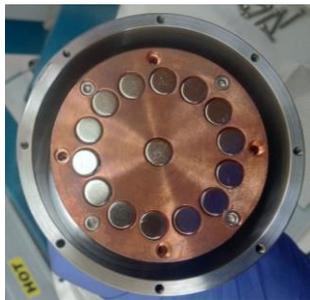
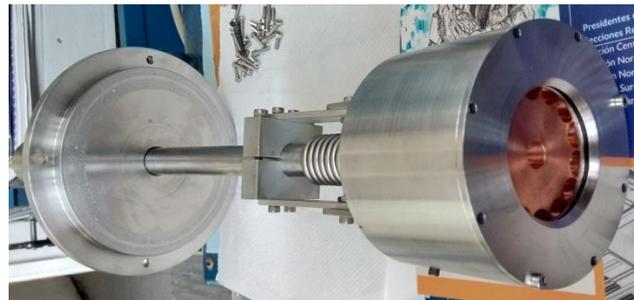

*Figura 5. Magnetrón sin tapa.*    *Figura 6. Magnetrón ensamblado en el brazo de la cámara de vacío.*

## 3. Resultados y discusión

La grafica de la figura 7 a) muestra el espesor de películas de Permalloy que se obtuvo para tiempos de depósito de 5, 10 y 15 minutos realizados con una potencia de 40W y una presión de $3 \times 10^{-3}$ Torr, el aumento en los espesores de las películas de permalloy respecto al tiempo muestran un comportamiento lineal, se realizó un ajuste lineal a estos datos con lo que se obtuvo una tasa de velocidad de depósito de 16.2 nm/min para el permalloy bajo estas condiciones de depósito. La grafica de la figura 7 b) muestra la variación en el espesor de una película de Permalloy en función de la distancia, tomando el cero en el centro del magnetrón. Se observa un espesor aproximado de 100 nm a una distancia de 0 cm (centro del magnetrón), este espesor va disminuyendo conforme nos alejamos del centro de depósito, llegando hasta un espesor de aproximado de 25 nm a una distancia de 7 cm. Aquí se puede observar que se mantiene un espesor similar del centro hasta el primer centímetro, esto permite asegurar que se puede colocar muestras en un diámetro de 2 cm y obtener un espesor homogéneo.

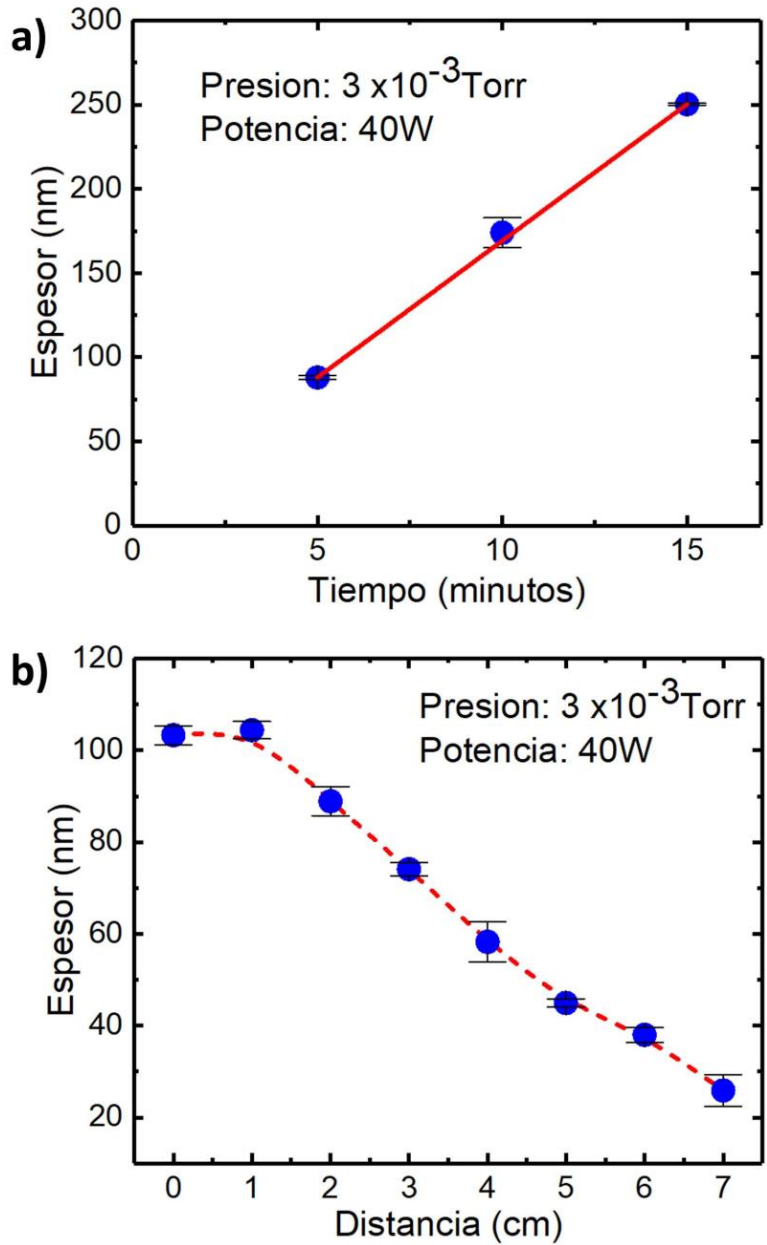

*Figura 7. a) Depósitos de Permalloy a una presión de 3x10⁻³Torr y una potencia de 40W, los círculos azules corresponden al espesor de la película obtenido para tiempos de 5, 10 y 15 minutos, la línea roja corresponde al ajuste lineal de los datos de los tres depósitos. b) Variación en el espesor de una película de Permalloy en función de la distancia, depositada a una presión de 3x10⁻³Torr y una potencia de 40W, los círculos azules representan el espesor de la película en diferentes distancias, la línea punteada roja es solo una guía visual. En ambas graficas las barras de error fueron obtenidas de tres diferentes medidas de espesor.*

La figura 8 muestra una micrografía tomada a un aumento de 50000x de la superficie de la película de Permalloy depositada a una presión de 3x10$^{-3}$ Torr y una potencia de 40W por un tiempo de 15 minutos, se puede observar una película continua de Permalloy sobre el sustrato de silicio. En la tabla 1 se muestran los datos del análisis elemental semicuantitativo obtenido por la técnica de EDS (Siglas del inglés, Scanning electron microscopy) en la película de Permalloy depositada a una presión de 3x10$^{-3}$ Torr y una potencia de 40W por un tiempo de 15 minutos. Se puede observar que los valores obtenidos en la película son muy similares a los valores estequiométricos del blanco de Permalloy (Ni$_{80}$Fe$_{20}$).

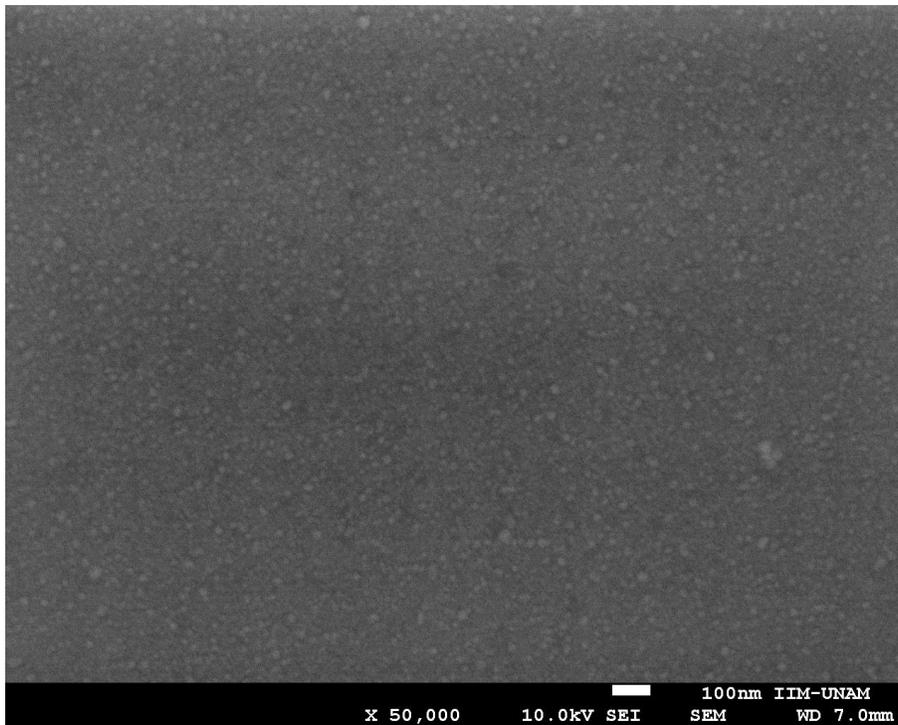

*Figura 8. Micrografía de una película de Permalloy depositada a una presión de 3x10$^{-3}$Torr y una potencia de 40W por un tiempo de 15 minutos.*

*Tabla 1. Análisis elemental semicuantitativo de una película de Permalloy depositada a una presión de 3x10$^{-3}$Torr y una potencia de 40W por un tiempo de 15 minutos.*

| Elemento | % Atómico |
|---|---|
| Ni | 80.27 |
| Fe | 19.73 |
| Total | 100 |

**4. Conclusión**

Se diseñó y fabrico un magnetrón de pulverización catódica para el uso de blancos ferromagnéticos de 2 pulgadas de diámetro. Se depositaron películas ferromagnéticas manométricas utilizando un blanco ferromagnético de Permalloy (Ni$_{80}$Fe$_{20}$). Se obtuvo la tasa de depósito para las películas delgadas de permalloy mediante un ajuste lineal de datos de tres diferentes espesores obtenidos de tres diferentes tiempos de depósito. Mediante microscopia electrónica de barrido se observó la formación de una película continua de permalloy sobre el sustrato y se obtuvo un porcentaje atómico de las películas similar a los porcentajes estequiométricos del blanco de Permalloy (Ni$_{80}$Fe$_{20}$). Este trabajo muestra que con un magnetrón fabricado en casa se pueden obtener películas delgadas de gran calidad, las cuales pueden ser utilizadas para la caracterización y estudio de una gran variedad de propiedades.

## 5. Agradecimientos